# A molecular dynamics study of path-dependent grain boundary properties in nanocrystals prepared using different methods

Hao Sun[a], Laurent Karim Béland[a,*]

[a]Department of Mechanical & Materials Engineering, Queen's University, 60 Union Street,

Kingston, Ontario, Canada

Grain boundaries are thermodynamically unstable. Hence, their properties should be path-dependent: grain boundaries in nanocrystals prepared using different methods might exhibit different properties. Using molecular dynamics simulations, we investigated grain boundaries of nanocrystals formed by quenching, solidification with pre-induced nucleation sites, and Voronoi tessellation. Some properties were found to be path-depend: the quenched model has lower boundary energy per atom, smaller boundary excess free volume per atom, and slower grain growth than the Voronoi model. We surmise that these differences are attributed to the abundant annealing twins in the quenched model. On the other hand, other properties are path-independent, such as Young's modulus, Poisson's ratio, and the ratio between grain boundary energy and boundary excess free volume. The results of this study further the understanding of the structural-property relationship of nanocrystals and provide guidance to future simulation-based studies of nanocrystalline materials.

* Corresponding author: laurent.beland@queensu.ca

Most metallic materials are polycrystals, whose 3D network of grain boundaries (GBs) influences their strength [1,2], ductility [3,4], and corrosion resistance [5]. These influences are magnified in nanocrystalline materials, owing to the large volume fraction of GBs present. In addition to their dependence on grain size, nanocrystalline properties also depend on GB structure. For example, replacing the original GBs with lower energy, more ordered coincident-site-lattice GBs can mitigate intergranular corrosion [6] and embrittlement [7–9]. GBs with larger excess free volumes have higher impurity segregation energy and therefore limit grain growth [10,11]. More generally, the structure-property relationship of GBs is tightly linked to restraining grain growth and improving the structural stability of nanocrystals, a topic of active research.

Unlike point defects, GBs are not thermodynamically stable. Accordingly, their structure and properties are path-dependent. In other words, GBs in nanocrystals created using different methods might exhibit different properties. While most polycrystals are formed via solidification of a liquid, nanocrystals can be fabricated without melting, such as by electron [12–14] or chemical vapor deposition [15], severe plastic deformation [16], and mechanical alloying [17]. In nanocrystals, GBs are formed when two growing crystal nuclei contact each other. In contrast, during solidification of traditional polycrystals, atoms between separate crystal nuclei exist in their liquid phase, and move collectively over long distances. Different formation procedures might lead to GB with different properties even if the average grain size and crystallographic orientation distribution were the same.

Molecular dynamics (MD) simulations allow the study of GBs with a high degree of precision. Two main methods are employed to prepare nanocrystalline MD models: Voronoi tessellation [10,18–20] and quenching [21–24]. The literature provides little guidance into which method is more representative of experimental procedures. One expects different preparation procedures to lead to GBs with different structures and properties, but—to the best of our knowledge—the literature contains no direct comparison. Specifically, due to high cooling rates ($> 10^{10} K/s$ [21–24]), nanocrystal models formed by quenching tend to contain large concentrations of solidification defects, such as annealing twins [21]. The effects of these annealing twins on GB structures and properties have hitherto remained unexplored.

In this study, molecular dynamics simulations were performed to investigate GB properties in nanocrystals formed by different methods, including direct quenching, solidification with pre-induced nucleation sites, and Voronoi tessellation. The GB properties being investigated in our study include GB energy, excess free volume, elastic moduli, and grain growth rate. We contrast the GB properties of simulated samples prepared using different methods, and explore the temperature and grain-size effects on GB properties.

MD simulations were performed using LAMMPS [25] and an embedded-atom-method potential for nickel [26], which involved properties of liquid nickel in its fitting procedure and correctly captures the formation energy of point defects. An accurate description of disordered structures and point defect energies is essential to simulate properties and phenomena related to GBs, which have high concentrations of point defects [27] and disordered structures at high temperatures [10]. As a compromise between computational limitation and physical correctness, we chose a timestep of 1 fs and set models to be 25 nm×25 nm×25 nm in $x$, $y$, and $z$ directions (Fig. 1(a)), accommodating approximately 1.3 million atoms. Atomic configurations were analyzed using common neighbor analysis with variable cut-off radius [28] as implemented in Ovito [29].

Different methods are used to create our nanocrystalline models. Two models involved quenching from the liquid phase at high cooling rates ($1.6 \times 10^{11}$K/s and $1.6 \times 10^{10}$K/s), and six models are solidified with different volume fractions of pre-induced nuclei. The shapes and orientations of these nuclei were taken from grains in a nanocrystalline model with an average grain size of 5 nm, created by Voronoi tessellation [30] (Fig. 1(a)). The dimensions of these nuclei were shrunk to leave space for liquid structures between different nuclei. The solidification and relaxation procedures were performed in the isothermal-isobaric ensemble (NPT), at 900 K and zero pressure. At last, four models were created using Voronoi tessellation [30] with different average grain sizes. After relaxation at zero pressure, zero temperature by conjugate gradient algorithm [31], these Voronoi models were gradually heated up to the target temperatures (at zero pressure) in 100 ps. The GB excess free volume (EFV) per atom is calculated as $\left(V_{\text{atom}}^{\text{GB}} - V_{\text{atom}}^{\text{crystal}}\right)/V_{\text{atom}}^{\text{crystal}}$, where $V_{\text{atom}}^{\text{GB}}$ and $V_{\text{atom}}^{\text{crystal}}$ are the atomic volume of GB atoms and crystal atoms, respectively. To measure GB moduli, we simulated strain-controlled, uniaxial tension of our models with a constant strain rate of $5 \times 10^8$/s, at different temperatures from 0 K to 1200 K. Young's modulus of GBs and grain interiors are calculated as the slope of the corresponding stress-strain curves at strains below 0.01.

Since nanocrystals are thermodynamically unstable, their structures are sensitive to their formation procedures. Three nanocrystalline models created by different methods are illustrated in Fig. 1(b). The model formed by direct quenching from liquid has a 3-nm average grain size and large amounts of annealing twins (Fig. 1(b)). Even after slowing down the cooling rate by a factor ten, the volume fraction of annealing twins barely changes (Fig. 1(c)). In contrast, solidification with pre-induced nuclei formed a nanocrystal similar to the Voronoi model from which these nuclei were taken (Fig. 1(b)). The larger volume fractions of these nuclei, the fewer annealing twins will remain (Fig. 1(c)). Among the three methods, the Voronoi-based model has the fewest annealing twins. Although the concentration of annealing twins in the Voronoi-based models increases with decreasing grain sizes, it is still much lower than that found in the quenched model even at the same average grain size (Fig. 1(d)). Given the much smaller cooling rates used in experiments ($< 10^9 K/s$ [32]), we believe the Voronoi-based model is more representative of experiments than the quenched model.

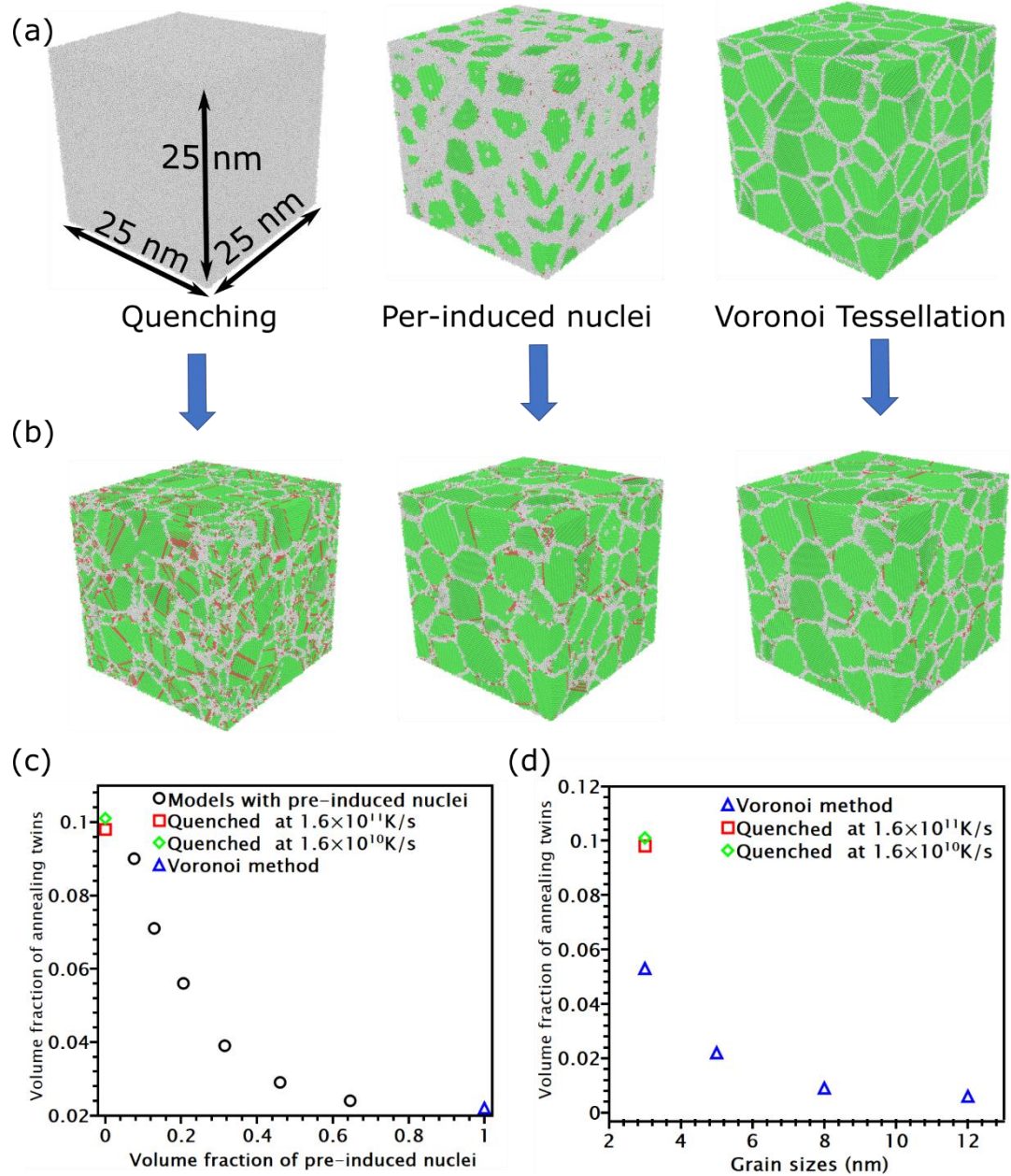

**Figure 1**. Nanocrystals formed by three methods: directly quenching from liquid, solidification with pre-induced nucleation nuclei, and Voronoi tessellation. The atomic configuration for the three models before and after solidification are illustrated in (a) and (b), respectively. Annealing twins are illustrated by red atoms; green and while atoms represent FCC and disordered structures, respectively. (c) shows the volume fraction of annealing twins in different models as a function of the volume fraction of pre-induced nuclei. Fractions equal to zero and one represent the quenched and Voronoi-based model, respectively. (d) exhibits the volume fraction of annealing twins as a function of the average grain sizes of Voronoi-based nanocrystalline models. The results from quenched models are also shown for comparison.

Analogously to the structural differences arising from different preparation procedures, GBs in different models might also exhibit different properties, such as different GB energy and EFV. Since grain growth during annealing decreases the total energy and releases the EFV, the total energy of our simulation models decreases linearly with the total volume (Fig. 2(a)). The slope of this linear relationship represents the ratio between GB energy and EFV, i.e., the GB energy increment per unit EFV. A constant value, $0.29\ eV/Å^3$, is found to be suitable for GBs in different models regardless of their formation procedures (Fig. 2(b)). GB energy has been found to follow a simple proportional correlation with the EFV [10,33–36]. Our proportionality constant, $0.29\ eV/Å^3$, is consistent with that for GBs in nickel calculated in a previous scientific study [35].

Despite sharing the same ratio between GB energy and EFV, GBs in different models could also have different energy and EFV per GB atom [10]. Since both energy and volume of our models decrease linearly with the remaining GB atoms, the slopes of the two linear relationships are the excess energy and EFV per GB atom, respectively. Due to the same energy increment per unit EFV (Fig. 2(b)), GB energies and EFV per atom in different models follow a linear relationship (Fig. 2(c)). The Voronoi model exhibits the highest energy and largest EFV and the quenched model shows the lowest. However, since grain growth only transforms a small portion of the total GBs atoms into crystalline structures (approximately 7% of at 900 K in 0.5 ns), the results shown in Fig. 2(c) only reflect local GB features. The average properties of all GB atoms can be calculated as the excess energy and EFV of the nanocrystalline model compared to that of a single-crystal model with the same number of atoms, divided by the total number of GB atoms. The results are shown in Fig. 2(d). Based on that alternative metric, the quenched model also possesses the lowest GB energy and EFV per atom. Furthermore, the average EFV and GB energy per atom at different temperatures follow a linear relationship (Fig. 2(c)) with the same slope. Thus, the ratio between GB energy and EFV, or the energy increment per EFV, is temperature-independent. Generally, our results indicate that the quenching process can transform a GB network into a lower energy state with smaller EFV, while the ratio between GB energy and EFV remains unchanged.

The GB structural relaxation and energy reduction in nanocrystals can be triggered by plastic deformation [37,38], realized via GB emission of stacking faults and twin boundaries. Specifically, shear-induced nanometer twins in nickel were observed to halve GB energy [39]. Likewise, annealing twins in the quenched model can replace a higher-energy GB with one having lower energy, an energy reduction unachievable without twinning [40]. GB energy reduction also diminishes EFV due to their proportional correlation. In addition, the quenched model has larger triple junctions than the Voronoi model (Fig. 2(f)) due to a "locking" of densities: during solidification, the remaining liquid phase at the meeting of three crystallization fronts has a lower density than the already crystallized phase [41]. This lack of atoms leads to larger triple junctions containing more disordered atoms that that in the Voronoi model. As EFV is necessary to accommodate the misfit across the boundary plane [42–44], larger triple junctions with more atoms diminish the geometrical misfit undertaken by each atom, thereby reducing the EFV and GB energy per atom. Accordingly, both annealing twins and larger triple junctions reduce GB energy and EFV per atom.

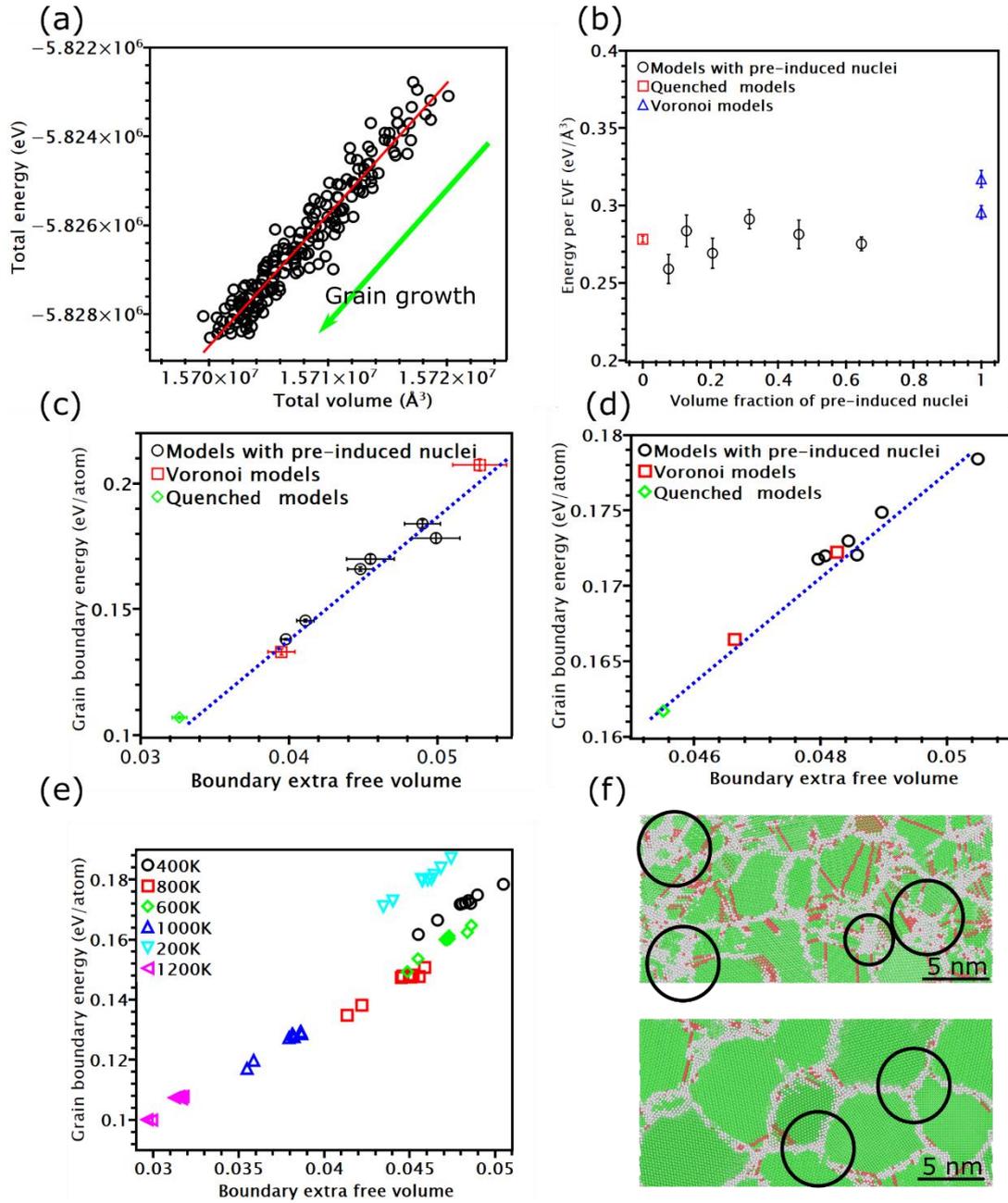

**Figure 2.** GB energy and excess free volume in different models. During grain growth, the total energy reduction follows a linear relationship with total volume shrinkage, as shown in (a). The slopes of this linear relationship for different models are illustrated in (b). (c) shows the energy and EFV per GB atom obtained from grain growth at 1000 K. (d) exhibits the energy and EFV per atom calculated from the excess energy of nanocrystals at 400 K. The results at different temperatures are exhibited in (e). (f) shows the atomic configurations at 0 K for models created by Quenching and Voronoi tessellation, respectively. Triple junctions are highlighted by black circles.

In addition to decrease GB energy, annealing twins and triple junctions also affect GB migration. Fig. 3(a) shows the remaining GB fractions during annealing at 800 K. While the GB fraction in the quenched model decreases linearly with time, that in the Voronoi model features a nonlinear drop in the initial 200 ps. In the quenched model, the annealing-twin fraction decreases linearly with time (Fig. 3(b)); in contrast, the annealing-twin fraction in the Voronoi model is approximately constant. This indicates that grain growth in the Voronoi model is mainly undertaken by the migration of GBs free of annealing twins, which suggests that annealing twins retard grain growth.

Using the slope in Fig. 3(a) of the linear regime as a representative grain growth rate, we generated an Arrhenius plot of the representative rates. The rate follows two distinct relations with $1/kT$, corresponding to two activation energies for GB migration. Such transitions are thought to be linked to a periodic-to-disordered GB structural transition [27,45–47]. GBs in the quenched model exhibits lower growth rate than that in the Voronoi model below 600K. In the quenched model, annealing twins left multiple triple junctions with adjoining GBs. Similarly to dislocations, triple junctions are line defects, whose migration energy barrier is analogous to Peierl's energy [48–50]. Only when this energy barrier is overcome by external force will the triple junction nucleate a new step in the direction of motion. Since the energy barrier for triple-junction migration is higher than that for GB migration [48,49], GB diffusion is hampered adjacent to triple junctions [51]. For example, in nanocrystal Cu, the sluggish nucleation of a new step of the triple junction between GBs and twin boundaries can slow down the overall atomic transport rate by one order of magnitude [51]. Thus, triple junctions are generally less mobile than GBs and can substantially retard GB migration. Nonetheless, the opposite is true above 900K: GBs in the quenched model become more mobile than that in the Voronoi model. Since triple junctions have higher migration energy barrier, their mobility is more sensitive to temperature than GB mobility [48,49]. Hence, triple junctions can become more mobile than GBs at high temperatures [48,49]. Rather than triple-junction retardation, the evolution of the GB network is rate-limited by GB migration.

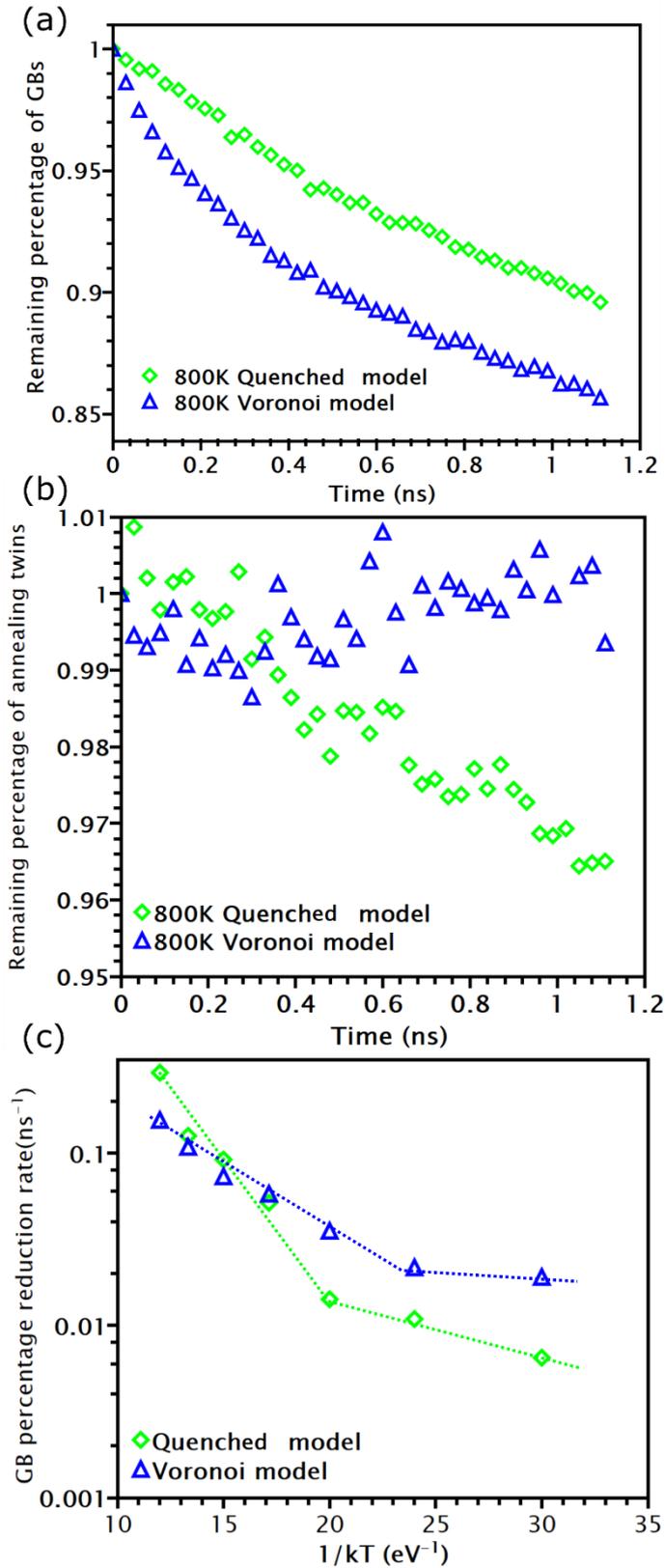

**Figure 3.** Grain growth retardation induced by annealing twins. (a) The remaining percentage of GBs during grain growth at 800K in a quenched model and a Voronoi model with the same average grain size. The remaining portion of annealing twins in these two models are exhibited in subfigure (b). (c) shows the representative grain growth rate as a function of $1/kT$.

We also investigated the Young's modulus and Poisson's ratio of different nanocrystals. Both are measured based on uniaxial tension with a constant strain rate of $5 \times 10^8/s$. The corresponding stress-strain curves for GBs, grain interior, and overall stress in a Voronoi's model at 0 K are illustrated in Fig. 4(a). The interaction between GBs and the grain interior leads to positive residual stresses in the GBs and negative stresses in the grain interior. Young's modulus of GBs is smaller than that of the grain interior's due to their EFV. The overall stress measured on the loading surface equals the weighed mean of GB stress and grain-interior stress based on their volume fraction (Fig. 4(a)). In other words, the mixture rule used in composites is also valid for nanocrystals. Note that our low-temperature simulations ignore quantum nuclear effects, which also affect mechanical properties [52].

Despite their difference in GB migration behaviours, the quenched model exhibit approximately the same Young's modulus as that of the Voronoi model having the same average grain size at different temperatures (Fig. 4(b) and (c)). Both the Young's moduli of GBs and grain interior decrease with increasing temperatures; Young's moduli of grain interior decrease faster than that of GBs. Increasing temperature enhances atomic thermal vibrations, which changes the atomic lattice's potential energy and the curvature of the potential energy curve, so Young's modulus will also change. It is experimentally established that Young's modulus follows a $T^4$ dependence at low temperatures and a $T$ dependence at high temperatures [53–55]. The $T^4$ dependence has been attributed to the third law of thermodynamics that the derivative of any elastic constant with respect to temperature must approach zero as the temperature approaches absolute zero [53–55]. The linear dependence is due to a high degree of excitation of vibrational modes when $kT \gg \hbar\omega_D$ ($\omega_D$ is the Debye frequency [56]) and classical statistical theory is valid. The relationship between Young's modulus and temperature, $dE/dT$, at high temperatures was found to be proportional to $\gamma(C_P/V)$, where $C_P$ is the specific heat at constant pressure, $V$ is the atomic volume, and $\gamma$ is Gruneisen's constant [57]. $\gamma$ reflects the volumetric dependence of phonon frequencies and increases with EFV [58]. Such increment is negligible when EFV is below 0.2 [59].

Because the EFV for all models are below 0.2 (Fig. 2(e)), GBs in our models have approximately the same $\gamma$ as the crystallites [60]. However, due to lower atomic densities, GBs have larger atomic volume $V$ and smaller $C_P$, so $dE/dT$ of GBs is smaller than that of grain interiors (Fig. 4(b) and (c)). Similar to Young's modulus, Poisson's ratio is also path-independent, following the same trend with temperature for different models. Models with smaller grain sizes have larger Poisson's ratios, consistent with a previous numerical study of nanocrystals [61].

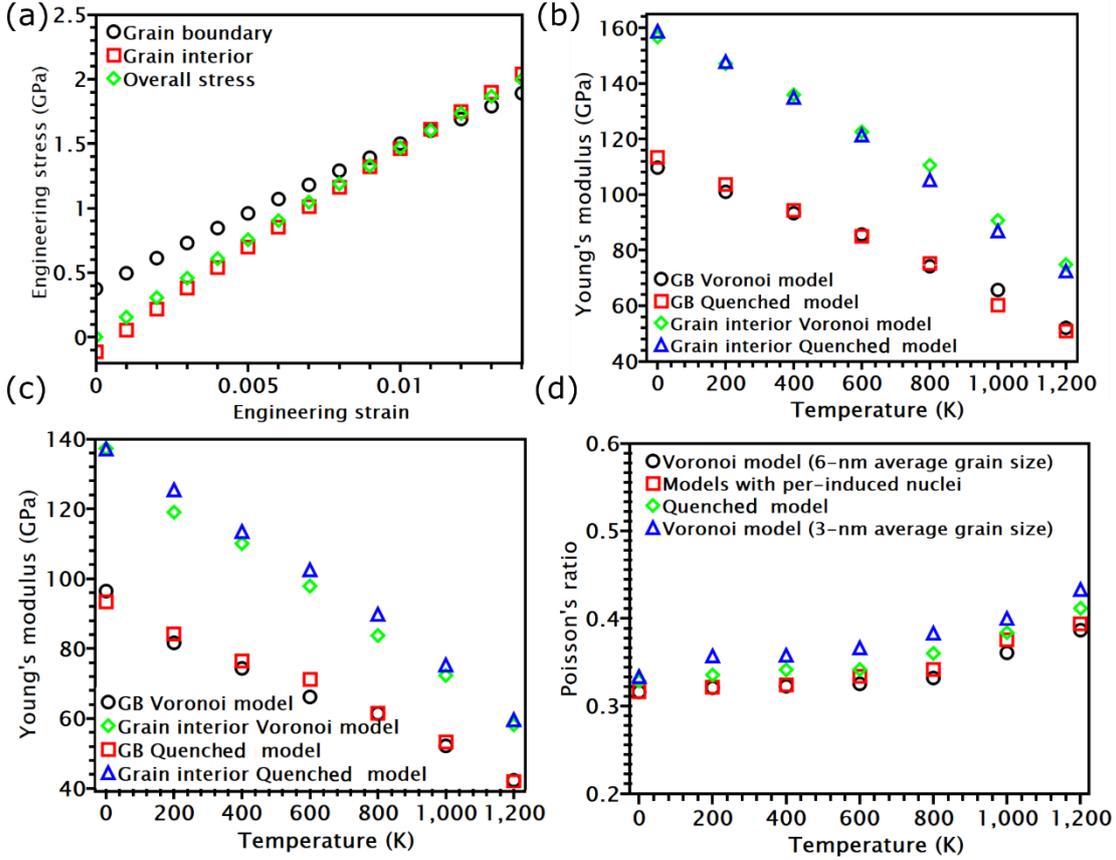

**Figure 4**. Young's modulus and Poisson's ratio of nanocrystals formed by different methods. (a) The stress-strain response of a Voronoi model with 6-nm average grain sizes. The overall stress, GB stress, and grain-interior stress are exhibited separately. (b) Young's modulus for Voronoi model and a model with pre-induced nuclei; both have an average grain size of 6 nm. (c) Young's modulus for Quenched model and Voronoi model; both have an average grain size of 3 nm. (d) The Poisson's ratio of different models at different temperatures.

To sum up, we investigated GB properties in nanocrystals prepared using direct quenching, solidification with pre-induced nucleation sites, and Voronoi tessellation. Some GB properties were found to be depend on the preparation procedure. Annealing twins in the quenched model lower GB energy per atom and diminish the EFV per atom compared to that in the Voronoi model. The triple junctions between GBs and annealing twins retard grain growth. On the other hand, some GB properties are path-independent, such as Young's modulus, Poisson's ratio, and the ratio between GB energy and EFV, which leads to a linear relationship between GB energy per atom and EFV per atom in all of the models considered in this study. The results of this study further the understanding of the structural-property relationship of nanocrystals and provide guidance to future simulation-based studies of nanocrystalline materials.

We thank Dr. Peyman Saidi for enlightening discussions. Financial support for this work was provided through the Natural Sciences and Engineering Research Council of Canada (NSERC). We also thank Compute Canada for generous allocation of computer resources.